\documentclass[aps,prd,preprint,superscriptaddress,showpacs,floatfix,nobibnotes]{revtex4}

\usepackage{latexsym}
\usepackage{amsmath}
\usepackage{amssymb}
\usepackage{graphicx}
\usepackage{longtable}

\usepackage{bm}
\usepackage{epsfig}
\usepackage{citesort}

\begin{document}
\title{Fourth-order Correlations of Conserved Charges in the QCD Thermodynamics}

\author{Wei-jie Fu}
\email[]{wjfu@itp.ac.cn} \affiliation{Kavli Institute for
Theoretical Physics China (KITPC), Key Laboratory of Frontiers in
Theoretical Physics, Institute of Theoretical Physics, Chinese
Academy of Science, Beijing 100190, China}

\author{Yue-liang Wu}
\email[]{ylwu@itp.ac.cn} \affiliation{Kavli Institute for
Theoretical Physics China (KITPC), Key Laboratory of Frontiers in
Theoretical Physics, Institute of Theoretical Physics, Chinese
Academy of Science, Beijing 100190, China}

\date{\today}

\begin{abstract}
The fourth-order correlations of conserved charges, such as the
baryon number, electric charge, and strangeness, are studied at
finite temperature and nonzero baryon chemical potential in an
effective model. It is found that the fourth-order correlations
change rapidly and have three extrema during the chiral crossover
with the increase of the baryon chemical potential. The absolute
values of the extrema approach infinity when the thermodynamical
system moves toward the QCD critical point and the fourth-order
correlations are divergent at the critical point. The contour plots
of the fourth-order correlations in the plane of the temperature and
baryon chemical potential are given. It is noticed that all the
fourth-order correlations of conserved charges, except for
$\chi_{13}^{BS}$ and $\chi_{13}^{QS}$, are excellent probes to
explore the QCD critical point in heavy ion collision experiments.
\end{abstract}


\maketitle

It is believed that the deconfined quark gluon plasma (QGP) is
produced in ultrarelativistic heavy ion
collisions~\cite{Shuryak2004,Gyulassy2005,Shuryak2005,Arsene2005,Back2005,Adams2005,Adcox2005,Blaizot2007}.
Therefore, QCD thermodynamics, such as the equation of state of the
QGP, chiral and deconfinement phase transitions, QCD phase diagram
and so on, has been a subject of intensive investigation in recent
years. One distinct characteristic of the QCD thermodynamics is that
there is a critical point in the QCD phase diagram in the plane of
temperature and baryon chemical potential, which separates the
first-order phase transition at high baryon chemical potential from
the continuous crossover at high temperature~\cite{Stephanov2006}.
This characteristic is confirmed in various field theory
models~\cite{Asakawa1989,Barducci1989,Barducci1994,Berges1999,Halasz1998,Scavenius2001,Hatta2003,Barducci2005,Fu2008}.
Although there is no definite evidence that the QCD critical point
also exists in the lattice QCD calculations due to the sign problem
at finite chemical potential, some lattice groups find that the QCD
critical point maybe exist in the phase diagram, based on
extrapolating results at small $\mu_{B}/T$ (ratio of the baryon
chemical potential and the temperature) to those at large
$\mu_{B}/T$~\cite{Fodor2002,Ejiri2004,Gavai2005}. In the meantime,
experiments with the goal to search for the QCD critical point are
planned and underway at the Relativistic Heavy Ion Collider (RHIC)
at the Brookhaven National Laboratory (BNL) and at the Super Proton
Synchrotron (SPS) at CERN in
Geneva~\cite{Mohanty2009,Anticic2009,Aggarwal2010,Aggarwal2010b}.

Then, searching for and locating the QCD critical point becomes a
crucial and vital task. It has been proposed that the QCD critical
point can be found through the non-monotonic behavior of
fluctuations and correlations of various particle multiplicities as
functions of varying control
parameters~\cite{Asakawa2000,Jeon2000,Stephanov1998,Stephanov1999,Hatta2003b,Jeon2004,Stephanov2009,Asakawa2009,Stephanov2010,Athanasiou2010,Fu2010}.
Particularly, fluctuations and correlations of conserved charges,
such as the baryon number, electric charge, and strangeness, deserve
more attentions. On the one hand, the fluctuations and correlations
of conserved charges are sensitive to the structure of the thermal
strongly interacting matter and behave differently between the
hadronic and QGP
phases~\cite{Asakawa2000,Jeon2000,Hatta2003,Jeon2004,Bhattacharyya2010}.
On the other hand, since the conserved charges are conserved through
the evolution of the fire ball, the fluctuations and correlations of
conserved charges can be measured in heavy ion collision
experiments.

In our previous work~\cite{Fu2010b}, we have calculated the
fluctuations of conserved charges up to the fourth-order and the
correlations to the third-order at finite temperature and nonzero
baryon chemical potential in the 2+1 flavor
Polyakov--Nambu--Jona-Lasinio (PNJL) model. We found an interesting
result: among all the fluctuations and correlations discussed in our
previous work~\cite{Fu2010b}, the numerical calculations indicate
that $\chi^{BQ}_{21}$, $\chi^{BS}_{21}$, $\chi^{QS}_{21}$, and
$\chi^{BQS}_{111}$ (those notations will be introduced in the
following) are the most valuable probes for exploring the QCD
critical point. All these quantities are the third-order
correlations of conserved charges. Therefore, it is natural to go
beyond our previous work to study the fourth-order correlations of
conserved charges near the QCD critical point. Furthermore, using
higher-order moments of particle multiplicity distributions to
search for the QCD critical point is possible in the Beam Energy
Scan at RHIC~\cite{Aggarwal2010}. In this work, we will calculate
the fourth-order correlations of conserved charges at finite
temperature and nonzero baryon chemical potential in the PNJL model.
Most attentions will be paid on the studies of the non-monotonic
behavior of the fourth-order correlations near the QCD critical
point.

We begin with the definition of the correlations of conserved
charges as follows
\begin{equation}
\chi_{ijk}^{BQS}=\frac{\partial^{i+j+k}(P/T^{4})}
{\partial(\mu_{B}/T)^{i}\partial(\mu_{Q}/T)^{j}\partial(\mu_{S}/T)^{k}},\label{susceptibility}
\end{equation}
where $P$ and $T$ is the pressure and temperature of a
thermodynamical system; $\mu_{B,Q,S}$ are the chemical potentials
for baryon number, electric charge, and strangeness, respectively.
These conserved charge chemical potentials are related with the
quark chemical potentials through the following relations,
\begin{equation}
\mu_{u}=\frac{1}{3}\mu_{B}+\frac{2}{3}\mu_{Q},\quad
\mu_{d}=\frac{1}{3}\mu_{B}-\frac{1}{3}\mu_{Q},\quad\textrm{and}\quad
\mu_{s}=\frac{1}{3}\mu_{B}-\frac{1}{3}\mu_{Q}-\mu_{S}.\label{chemicalpotential}
\end{equation}
where $\mu_{u,d,s}$ are the chemical potentials for $u$, $d$, and
$s$ quarks, respectively. Denoting the ensemble average of conserved
charge number $N_{X}$ ($X=B,Q,S$) with $\langle N_{X}\rangle$, we
can obtain the fourth-order correlations as follow
\begin{eqnarray}
\chi_{13}^{XY}&=&\frac{1}{VT^{3}}\Big(\langle \delta N_{X} {\delta N_{Y}}^{3}\rangle
-3\langle \delta N_{X}\delta N_{Y}\rangle\langle{\delta N_{Y}}^{2}\rangle\Big),  \\
\chi_{22}^{XY}&=&\frac{1}{VT^{3}}\Big(\langle {\delta
N_{X}}^{2}{\delta N_{Y}}^{2}\rangle -\langle{\delta
N_{X}}^{2}\rangle\langle{\delta N_{Y}}^{2}\rangle \nonumber\\
&&-2{\langle\delta N_{X}\delta N_{Y}\rangle}^{2}\Big),\\
\chi_{211}^{XYZ}&=&\frac{1}{VT^{3}}\Big(\langle {\delta N_{X}}^{2}
\delta N_{Y} \delta N_{Z}\rangle-2\langle\delta N_{X}\delta
N_{Y}\rangle\langle\delta N_{X}\delta N_{Z}\rangle\nonumber\\
&&-\langle{\delta N_{X}}^{2}\rangle\langle\delta N_{Y}\delta
N_{Z}\rangle\Big),
\end{eqnarray}
where $\delta N_{X}\equiv N_{X}-\langle N_{X}\rangle$ and $V$ is the
volume of the system.

We adopt the 2+1 flavor Polyakov-loop improved NJL model to study
the fourth-order correlations of conserved charges near the QCD
critical point. The validity of this effective model is expected,
since the critical behavior of the QCD phase transition is governed
by the universality class of the chiral symmetry, which is kept in
this model. Furthermore, compared with the conventional
Nambu--Jona-Lasinio model, the PNJL model not only has the chiral
symmetry and its dynamical breaking mechanism, but also includes the
effect of color confinement through the Polyakov
loop~\cite{Meisinger9602,Pisarski2000,Fukushima2004,Ratti2006a,Ratti2006b,Ciminale2008,Fu2008,Fu2009,Ghosh2006}.
Furthermore, the fluctuations and correlations of conserved charges
calculated in the 2+1 flavor PNJL model at finite temperature but
with vanishing chemical potentials in Ref.~\cite{Fu2010} are well
consistent with those obtained in lattice
calculations~\cite{Cheng2009}, which shows that the 2+1 flavor PNJL
model is well applicable to study the cumulants of conserved charge
multiplicity distributions.

The Lagrangian density for the 2+1 flavor PNJL model is given
as~\cite{Fu2008}
\begin{eqnarray}
\mathcal{L}_{\mathrm{PNJL}}&=&\bar{\psi}(i\gamma_{\mu}D^{\mu}+\gamma_{0}
 \hat{\mu}-\hat{m}_{0})\psi
 +G\sum_{a=0}^{8}\Big[(\bar{\psi}\tau_{a}\psi)^{2}
 +(\bar{\psi}i\gamma_{5}\tau_{a}\psi)^{2}\Big]   \nonumber \\
&&-K\Big[\textrm{det}_{f}(\bar{\psi}(1+\gamma_{5})\psi)
 +\textrm{det}_{f}(\bar{\psi}(1-\gamma_{5})\psi)\Big]
 -\mathcal{U}(\Phi,\Phi^{*} \, ,T),\label{lagragian}
\end{eqnarray}
where $\psi=(\psi_{u},\psi_{d},\psi_{s})^{T}$ is the three-flavor
quark field, and
\begin{equation}
D^{\mu}=\partial^{\mu}-iA^{\mu}\quad\textrm{with}\quad
A^{\mu}=\delta^{\mu}_{0}A^{0}\quad\textrm{,}\quad
A^{0}=g\mathcal{A}^{0}_{a}\frac{\lambda_{a}}{2}=-iA_4,
\end{equation}
where $\lambda_{a}$'s are the Gell-Mann matrices in color space and
$g$ is the gauge coupling strength.
$\hat{m}_{0}=\textrm{diag}(m_{0}^{u},m_{0}^{d},m_{0}^{s})$ is the
three-flavor current quark mass matrix. Throughout this work, we
take $m_{0}^{u}=m_{0}^{d}\equiv m_{0}^{l}$, while keep $m_{0}^{s}$
being larger than $m_{0}^{l}$, which breaks the $SU(3)_f$ symmetry.
The matrix $\hat{\mu}=\textrm{diag}(\mu_{u}, \mu_{d}, \mu_{s})$
denotes the quark chemical potentials which are related with the
conserved charge chemical potentials through relations in
Eq.(\ref{chemicalpotential}).

In the above PNJL Lagrangian,
$\mathcal{U}\left(\Phi,\Phi^{*},T\right)$ is the Polyakov-loop
effective potential, which is expressed in terms of the traced
Polyakov-loop $\Phi=(\mathrm{Tr}_{c}L)/N_{c}$ and its conjugate
$\Phi^{*}=(\mathrm{Tr}_{c}L^{\dag})/N_{c}$ with the Polyakov-loop
$L$ being a matrix in color space given explicitly by
\begin{equation}
L(\vec{x})=\mathcal{P}\exp\Big[i\int_{0}^{\beta}d\tau\,
A_{4}(\vec{x},\tau)\Big] =\exp[i \beta A_{4}],
\end{equation}
with  $\beta=1/T$ being the inverse of temperature and
$A_{4}=iA^{0}$.

In our work, we use the Polyakov-loop effective potential which is a
polynomial in $\Phi$ and $\Phi^{*}$~\cite{Ratti2006a}, given by

\begin{equation}
\frac{\mathcal{U}(\Phi,\Phi^{*},T)}{T^{4}} =
-\frac{b_{2}(T)}{2}\Phi^{*}\Phi -\frac{b_{3}}{6}
(\Phi^{3}+{\Phi^{*}}^{3})+\frac{b_{4}}{4}(\Phi^{*}\Phi)^{2} \, ,
\end{equation}
with
\begin{equation}
b_{2}(T)=a_{0}+a_{1}(\frac{T_{0}}{T})+a_{2} {(\frac{T_{0}}{T})}^{2}
+a_{3}{(\frac{T_{0}}{T})}^{3}.
\end{equation}
The parameters in the effective potential are fitted to reproduce
the thermodynamical behavior of the pure-gauge QCD obtained from the
lattice simulations, and their values are $a_{0}=6.75$,
$a_{1}=-1.95$, $a_{2}=2.625$, $a_{3}=-7.44$, $b_{3}=0.75$, and
$b_{4}=7.5$. The parameter $T_{0}$ is the critical temperature for
the deconfinement phase transition to take place in pure-gauge QCD
and $T_{0}$ is chosen to be $270\,\mathrm{MeV}$ according to the
lattice calculations. As for the five parameters in the quark sector
of the model, we adopt their values which are
$m_{0}^{l}=5.5\;\mathrm{MeV}$, $m_{0}^{s}=140.7\;\mathrm{MeV}$,
$G\Lambda^{2}=1.835$, $K\Lambda^{5}=12.36$, and
$\Lambda=602.3\;\mathrm{MeV}$, which are fixed by fitting the
observables $m_{\pi}=135.0\;\mathrm{MeV}$,
$m_{K}=497.7\;\mathrm{MeV}$,
$m_{\eta^{\prime}}=957.8\;\mathrm{MeV}$, and
$f_{\pi}=92.4\;\mathrm{MeV}$~\cite{Rehberg1996}.

In the mean field approximation, the thermodynamical potential
density ($\Omega=-P$) for the 2+1 flavor quark system is given by
\begin{eqnarray}
\Omega&=&-2N_{c}\sum_{f=u,d,s}\int\frac{d^{3}p}{(2\pi)^{3}}\Big\{E_{p}^{f}\theta(\Lambda^{2}-p^{2})\nonumber \\
&&+ \frac{T}{3}\ln\big[1+3\Phi^{*}e^{-(E_{p}^{f}-\mu_{f})/T}+3\Phi
e^{-2(E_{p}^{f}-\mu_{f})/T}+e^{-3(E_{p}^{f}-\mu_{f})/T}\big]
\nonumber \\
&&+ \frac{T}{3}\ln\big[1+3\Phi e^{-(E_{p}^{f}+\mu_{f})/T}+3\Phi^{*}
e^{-2(E_{p}^{f}+\mu_{f})/T}+e^{-3(E_{p}^{f}+\mu_{f})/T}\big]\Big\}\nonumber \\
&&+2G({\phi_{u}}^{2}
+{\phi_{d}}^{2}+{\phi_{s}}^{2})-4K\phi_{u}\,\phi_{d}\,\phi_{s}+\mathcal{U}(\Phi,\Phi^{*},T),\label{thermopotential}
\end{eqnarray}

where $\phi_{i}$'s $(i=u,d,s)$ are the quark chiral condensates, and
the energy-momentum dispersion relation is
$E_{p}^{i}=\sqrt{p^{2}+M_{i}^{2}}$, with the constituent mass being
\begin{equation}
M_{i}=m_{0}^{i}-4G\phi_{i}+2K\phi_{j}\,\phi_{k}.\label{constituentmass}
\end{equation}
Minimizing the thermodynamical potential in
Eq.~\eqref{thermopotential} with respective to $\phi_{u}$,
$\phi_{d}$, $\phi_{s}$, $\Phi$, and $\Phi^{*}$, we obtain a set of
equations for the minimal conditions, which can be solved as
functions of temperature $T$ and three conserved charge chemical
potentials $\mu_{B}$, $\mu_{Q}$, and $\mu_{S}$.

\begin{figure}[!htb]
\includegraphics[scale=1.0]{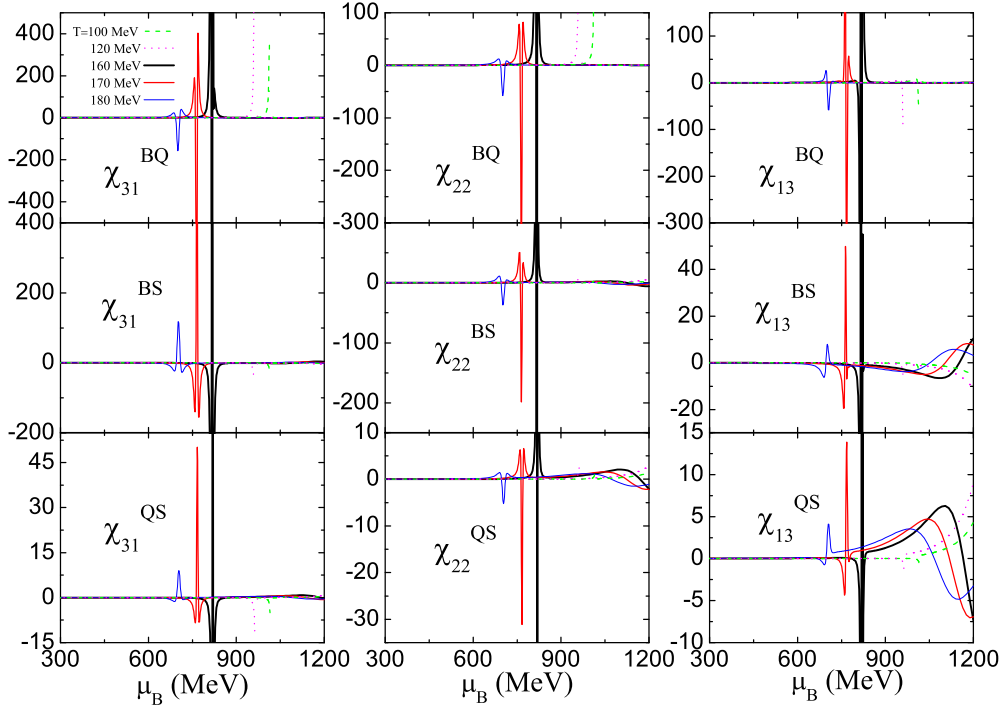}
\caption{(color online). Fourth-order correlations $\chi_{31}^{BQ}$
(top-left), $\chi_{22}^{BQ}$ (top-middle), $\chi_{13}^{BQ}$
(top-right), $\chi_{31}^{BS}$ (center-left), $\chi_{22}^{BS}$
(center-middle), $\chi_{13}^{BS}$ (center-right), $\chi_{31}^{QS}$
(bottom-left), $\chi_{22}^{QS}$ (bottom-middle), and
$\chi_{13}^{QS}$ (bottom-right) as functions of the baryon chemical
potential $\mu_{B}$ ($\mu_{Q}=\mu_{S}=0$) with several values of
temperature in the PNJL model.}\label{f1}
\end{figure}

We use the method of Taylor expansion to compute the fourth-order
correlations of conserved charges in the PNJL model. In
Fig.~\ref{f1} we show the fourth-order correlations between two
conserved charges versus the baryon chemical potential at several
values of temperature calculated in the PNJL model. We find that the
QCD critical point is located at about $T_{c}=160\;\mathrm{MeV}$ and
${\mu_{B}}_{c}=819\;\mathrm{MeV}$ ($\mu_{Q}=\mu_{S}=0$) with input
parameters given above. From Fig.~\ref{f1}, one can clearly see that
the magnitudes of all the fourth-order correlations in this figure
grow rapidly and oscillate drastically when the QCD phase transition
occurs. Comparing different curves corresponding to different
temperatures, we can recognize that the oscillating amplitudes of
the correlations increase quickly when moving toward the QCD
critical point, i.e., the temperature approaches to
$T_{c}=160\;\mathrm{MeV}$. All the fourth-order correlations between
two conserved charges diverge at the QCD critical point. When the
temperature is below $T_{c}$, the chiral phase transition is
first-order and the correlations are discontinuous during the phase
transition as shown by the dashed and dotted lines in Fig.~\ref{f1}.
When the temperature is above $T_{c}$, the chiral phase transition
is a continuous crossover due to finite quark current mass.
Correspondingly, the correlations of conserved charges are also
continuous during the QCD phase transition. As for the fourth-order
correlations between two conserved charges, one notices that there
are two maxima and one minimum in the curves of $\chi_{31}^{BQ}$,
$\chi_{22}^{BQ}$, $\chi_{13}^{BQ}$, $\chi_{22}^{BS}$, and
$\chi_{22}^{QS}$, while there are two minima and one maximum in the
curves of other fourth-order correlations between two conserved
charges. Furthermore, it is seen that all the fourth-order
correlations except for $\chi_{13}^{BS}$ and $\chi_{13}^{QS}$ vanish
rapidly once the thermodynamical system deviates from the QCD phase
transition. In Fig.~\ref{f1} we observe that $\chi_{13}^{BS}$ and
$\chi_{13}^{QS}$ still have finite values at large baryon chemical
potential, this is because the strange quark have relatively large
current mass and its constituent mass can not be neglected even in
the chiral symmetric phase.

\begin{figure}[!htb]
\includegraphics[scale=1.2]{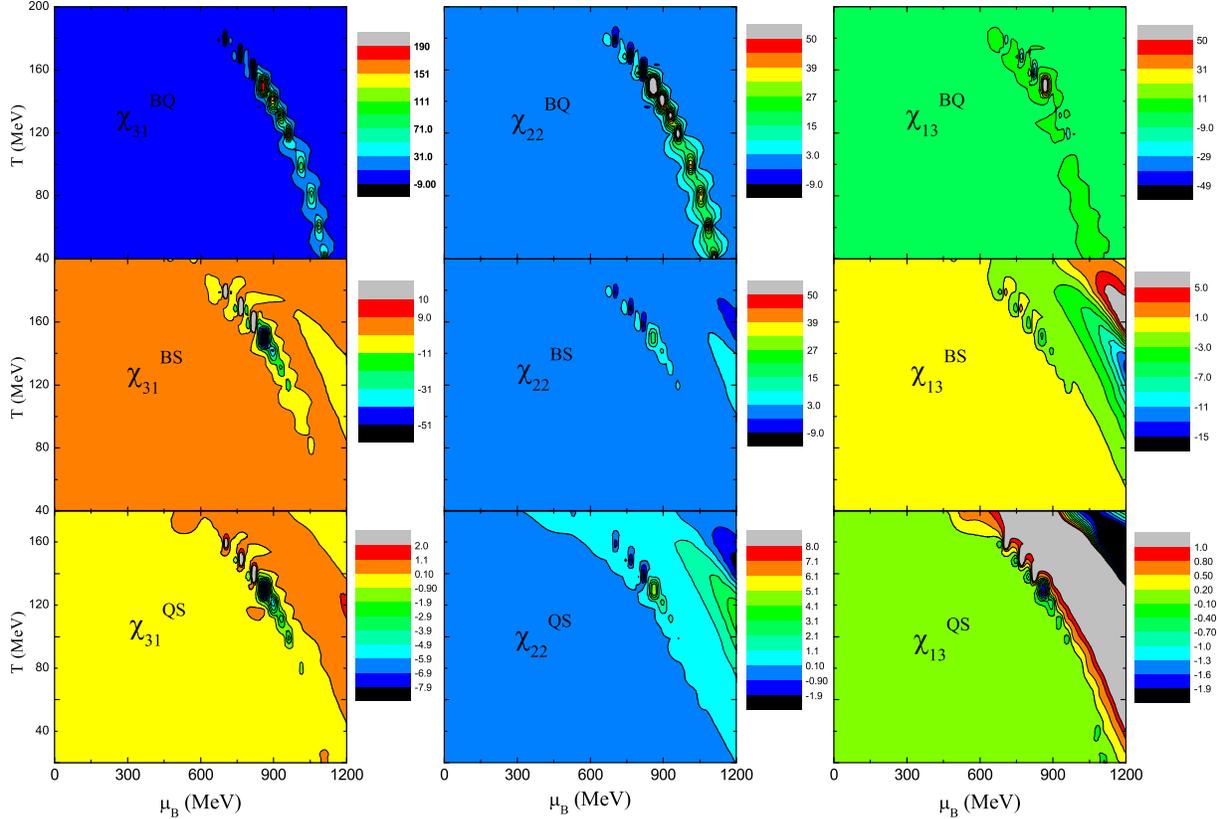}
\caption{(color online). Contour plots of the fourth-order
correlations $\chi_{31}^{BQ}$ (top-left), $\chi_{22}^{BQ}$
(top-middle), $\chi_{13}^{BQ}$ (top-right), $\chi_{31}^{BS}$
(center-left), $\chi_{22}^{BS}$ (center-middle), $\chi_{13}^{BS}$
(center-right), $\chi_{31}^{QS}$ (bottom-left), $\chi_{22}^{QS}$
(bottom-middle), and $\chi_{13}^{QS}$ (bottom-right) versus
temperature $T$ and baryon chemical potential $\mu_{B}$
($\mu_{Q}=\mu_{S}=0$) in the PNJL model.}\label{f2}
\end{figure}

\begin{figure}[!htb]
\includegraphics[scale=0.7]{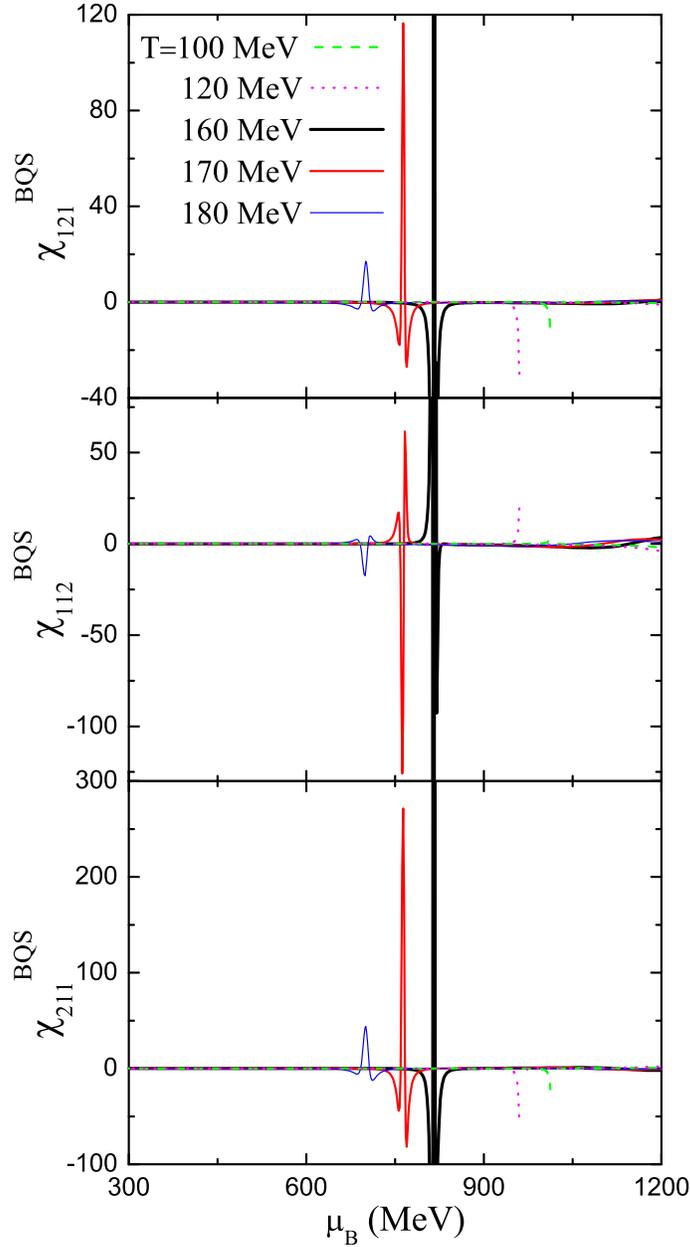}
\caption{(color online). Fourth-order correlations among three
conserved charges $\chi_{121}^{BQS}$ (top), $\chi_{112}^{BQS}$
(middle), and $\chi_{211}^{BQS}$ (bottom) as functions of the baryon
chemical potential $\mu_{B}$ ($\mu_{Q}=\mu_{S}=0$) with several
values of temperature in the PNJL model.}\label{f3}
\end{figure}

Fig.~\ref{f2} shows the contour plots of the fourth-order
correlations between two conserved charges as functions of the
temperature and baryon chemical potential calculated in the PNJL
model. We observe that the fourth-order correlations between two
conserved charges are vanishing in the chiral symmetry broken phase,
i.e., in the bottom-left region of every little plot. It is noticed
that the chiral phase transition line is distinct in these plots and
one can easily recognize the region near around the QCD critical
point, where the contour lines are dense. Comparing all these
fourth-order correlations between two conserved charges, we notice
that $\chi_{31}^{BQ}$, $\chi_{22}^{BQ}$, $\chi_{13}^{BQ}$,
$\chi_{31}^{BS}$, $\chi_{22}^{BS}$, $\chi_{31}^{QS}$, and
$\chi_{22}^{QS}$ are superior to $\chi_{13}^{BS}$ and
$\chi_{13}^{QS}$ to be used to search for the critical point in
heavy ion collision experiments, since the former seven correlations
have finite values only when the thermodynamical system is near
around the QCD critical point.

\begin{figure}[!htb]
\includegraphics[scale=0.8]{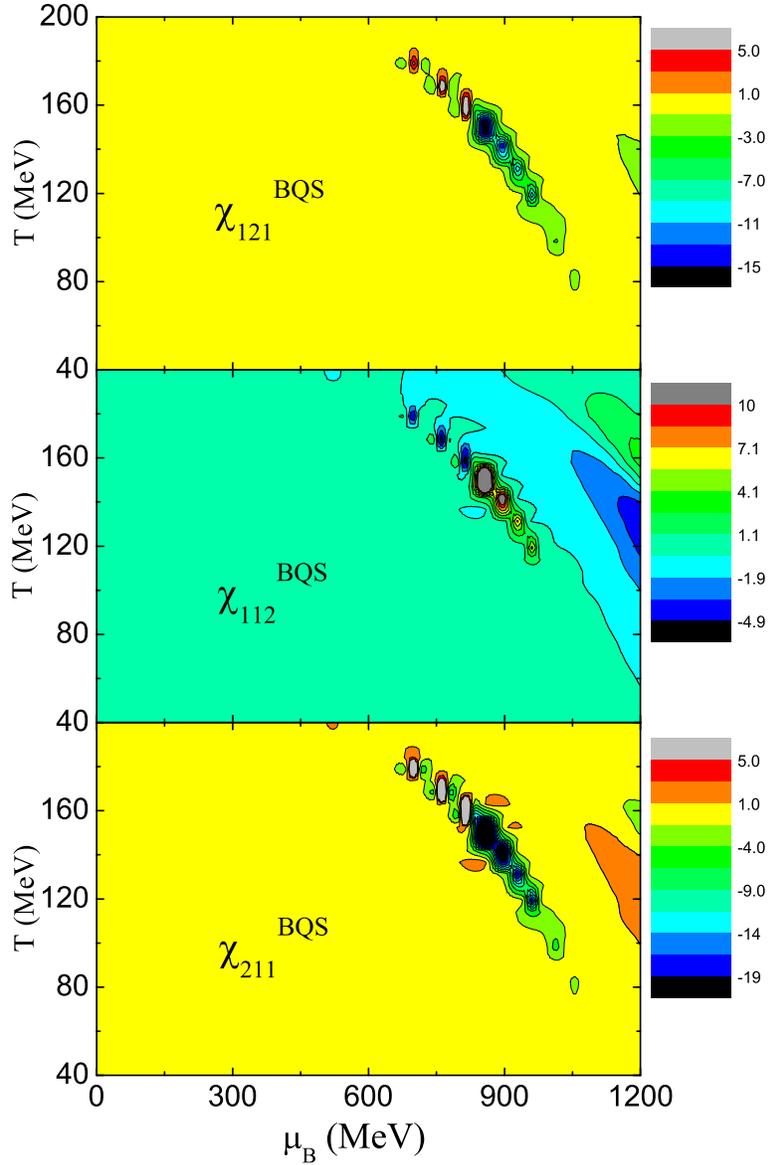}
\caption{(color online). Contour plots of the fourth-order
correlations among three conserved charges $\chi_{121}^{BQS}$ (top),
$\chi_{112}^{BQS}$ (middle), and $\chi_{211}^{BQS}$ (bottom) versus
temperature $T$ and baryon chemical potential $\mu_{B}$
($\mu_{Q}=\mu_{S}=0$) in the PNJL model.}\label{f4}
\end{figure}

In Fig.~\ref{f3} we present the fourth-order correlations among
three conserved charges as functions of the baryon chemical
potential with several values of temperature calculated in the PNJL
model. It is seen that the fourth-order correlations among three
conserved charges are divergent at the QCD critical point same as
those between two conserved charges. When the temperature is below
$T_{c}=160\;\mathrm{MeV}$, those correlations are discontinuous at
the first-order chiral phase transition; while when the temperature
is above $T_{c}$, they are continuous functions of the baryon
chemical potential and change rapidly at the chiral crossover. It is
observed that there are two minima and one maximum in the curves of
$\chi_{121}^{BQS}$ and $\chi_{211}^{BQS}$, and two maxima and one
minimum in the curve of $\chi_{112}^{BQS}$. One can also clearly
notice that the fourth-order correlations among three conserved
charges approach zero rapidly once the thermodynamical system
deviates from the chiral phase transition.

In Fig.~\ref{f4} we show the contour plots of the fourth-order
correlations among three conserved charges as functions of the
temperature and baryon chemical potential in the PNJL model. One can
clearly recognize the QCD critical point in these three plots, which
demonstrates that the fourth-order correlations among three
conserved charges are quite sensitive to the singular structure
related to the critical point. Therefore, the fourth-order
correlations among three conserved charges are excellent probes to
explore the QCD critical point in heavy ion collision experiments.

In summary, we have studied the fourth-order correlations of
conserved charges, such as the baryon number, electric charge, and
strangeness, at finite temperature and nonzero baryon chemical
potential in the 2+1 favor Polyakov-loop improved
Nambu--Jona-Lasinio model. It is found that the fourth-order
correlations of conserved charges are divergent at the QCD critical
point. When the temperature is below the critical temperature of the
QCD critical point, the fourth-order correlations are discontinuous
at the first-order chiral phase transition; while when the
temperature is above the critical temperature, the fourth-order
correlations of conserved charges are continuous with the change of
the baryon chemical potential and oscillate rapidly at the chiral
crossover. In the curves of the fourth-order correlations as
functions of the baryon chemical potential, there are three extrema
during the chiral crossover. Furthermore, we have given the contour
plots of the fourth-order correlations of conserved charges in the
plane of temperature and baryon chemical potential. The QCD critical
point can be easily recognized in these plots. Comparing the
fourth-order correlations, we notice that all the fourth-order
correlations of conserved charges, except for $\chi_{13}^{BS}$ and
$\chi_{13}^{QS}$, are excellent probes to explore the QCD critical
point, since these fourth-order correlations approach zero rapidly
once the thermodynamical system deviates from the QCD critical
point.

W. J. F. acknowledges financial support from China Postdoctoral
Science Foundation No. 20090460534. Y. L. W. is supported in part by
the National Science Foundation of China (NSFC) under the grant No.
10821504.

\end{document}